\documentclass[aps,twocolumn,showpacs,superscriptaddress,citeautoscript,reprint,prr,floatfix]{revtex4-2}
\usepackage{amsmath}
\usepackage{amssymb}
\usepackage{graphicx}
\usepackage[dvipsnames,svgnames]{xcolor}
\usepackage{mathtools}
\usepackage{bm}
\usepackage{siunitx}
\usepackage{braket}
\usepackage{gensymb}
\usepackage{colortbl}
\usepackage{booktabs}

\usepackage{soul}

\newcommand{\changed}{\textcolor{black}}
\definecolor{electricgreen}{rgb}{0.0, 1.0, 0.0}
\definecolor{babyblueeyes}{rgb}{0.63, 0.79, 0.95}

\begin{document}
\title{\changed{Fractional corner charges in threefold-symmetric two-dimensional materials with fragile topology}}
\author{Olga Arroyo-Gasc\'on}
\affiliation{\changed{Nanotechnology Group, USAL-Nanolab, University of Salamanca, E-37008 Salamanca, Spain}}
\author{Sergio Bravo}
\affiliation{Departamento de F\'isica, Universidad T\'ecnica Federico Santa Mar\'ia, Casilla 110-V, Valpara\'iso, Chile}
\author{Leonor Chico}
\affiliation{GISC, Departamento de F\'{\i}sica de Materiales, Facultad de Ciencias Físicas, Universidad Complutense de Madrid, E-28040 Madrid, Spain}
\author{Mónica Pacheco}
\affiliation{Departamento de F\'isica, Universidad T\'ecnica Federico Santa Mar\'ia, Casilla 110-V, Valpara\'iso, Chile}

\date{\today}

\begin{abstract}
We perform a systematic study of the signatures of fragile topology in 
over 50 nonmagnetic
two-dimensional
materials with formula AB$_2$,
belonging to space group $P\Bar{3}m1$. 
Using group theory analysis in the framework
of topological quantum chemistry, we find fragile bands near the Fermi level for all the materials studied. 
Since stable topological bands are also present in these systems, the interplay of both phases is discussed, showing that corner charges appear in 
over 80\% of the materials and \changed{are linked to} fragile topology. Using first-principles calculations, we predict fractionally-charged corner charges protected by $C_3$ symmetry. Our work aims to broaden the scope of materials with experimentally accessible fragile bands.
\end{abstract}

\maketitle

\section{Introduction}

Since the proposal and subsequent discovery of the first topological insulators \cite{Hasan2010,Qi2011}, topology has
enriched the electronic properties of a great number of materials. Different types of Dirac and Weyl fermions have been discovered in analogy to high-energy physics \cite{Wehling2014,Xu2015, Armitage2018}.
The development of symmetry indicators and topological quantum chemistry (TQC)
\changed{\cite{Fu2007,po_symmetry-based_2017,PRX_7,bradlyn_topological_2017-1,elcoro_application_2020,po_symmetry_2020,cano_band_2021,vergniory_all_2022}}
has further extended and standardized the variety of topological
phases of matter. Higher-order topological insulators (HOTIs) and fragile topology
are the latest additions; the former was first established in topological insulators
characterized with a mirror Chern number \cite{Teo2008,schindler_higher-order_2018}, while the
latter has been theoretically proposed in twisted bilayer graphene \changed{\cite{po_PhysRevLett121,zou_PhysRevB.98.085435,song_all_2019,po_faithful_2019,ahn_failure_2019}}. Both phases share the absence
of topological gapless modes on their surfaces or edges, in contrast to standard
topological insulators. Whereas $d-$ dimensional $n^{\text{th}}$-order topological insulators are known to host topological
states in \changed{$d-n$} dimensions, the same does not always apply to fragile topology.
Interestingly, while nearly 90\% of all classified bulk materials have at least one topological band 
\cite{vergniory_all_2022,wieder2022},
only about 8\% show fragile states. Thus, the search for
new materials that could provide more information on signals of fragile topology
is highly desirable.

Fragile topology is based on the fact that a set of fragile bands can be trivialized
by adding a suitable set of trivial bands \cite{po_PhysRevLett121,zou_PhysRevB.98.085435}. The starting point for every analysis of symmetry-indicated topology is to characterize the form in which the electronic bands transform at high symmetry points in reciprocal space, under the action of the space group of the system. This allows to classify each band 
using the eigenvalues of the symmetry operations of the material, which corresponds to the characters (traces) of the irreducible representations (irreps). Due to real space periodicity, a set of orbitals located at specific Wyckoff positions (WP) will induce a well-defined set of irreps 
in the momentum space description of the bands. The key point is that in the atomic limit, that is, no interaction among the orbitals is considered, the orbitals will induce what is called an elementary band representation (EBR). These EBRs are the building blocks that allow discriminating the type of band topology in a material \cite{bradlyn_topological_2017-1,elcoro_application_2020,cano_band_2021}. Thus, a band is said to be topologically trivial if it can be expressed as a linear combination of EBRs where all coefficients are positive integers \cite{elcoro_application_2020,cano_band_2021}. Unlike trivial bands, fragile bands can be
expressed as a combination of EBRs with integer coefficients, but at least one of them
must be negative \cite{elcoro_application_2020}. \changed{This way, when more bands are added
to a fragile one, the negative EBRs can cancel out, giving rise to an all-positive
linear combination of EBRs and thus trivializing the band set \cite{PRX_7,bradlyn_topological_2017-1,po_PhysRevLett121,song_all_2019,cano_band_2021,vergniory_all_2022,wieder2022}. Hence, even if the
fragile state is near the Fermi energy, the lower energy bands can trivialize it.
This trivialization can explain the scarcity of topological materials with fragile
topology features. For its part, a band has stable topology (stable upon band addition) if it can be written in terms of EBRs and at least one coefficient is fractional \cite{PRX_7,bradlyn_topological_2017-1,elcoro_application_2020,cano_band_2021,vergniory_all_2022,wieder2022}. }

From the above discussion, fragile topology cannot be captured by stable
topological invariants such as the $\mathbb{Z}_2$ index used in spin Hall
systems. Thereby, TQC stands as one of the most useful tools to diagnose 
this kind of topology, together with the most general study of the evolution
of Wannier charge centers along appropriate directions in reciprocal space   
\cite{benalcazar_electric,cano_topology_2018,song_all_2019,bradlyn_disconnected_2019,bouhon_PhysRevB.100.195135,twisted_fragile,bouhon_fragile}.
\changed{Beyond TQC, stable topological phases were already established in relation to the tenfold classification of systems attending to their (non-crystalline) symmetries \cite{altland_nonstandard_1997} and within K-theory \cite{Kitaev2009}. Note that, in a similar way, fragile phases can also arise without crystalline protection and even beyond symmetry indicators \cite{bouhon_fragile}, as is also the case for other recently-discovered topological phases \cite{wang_higher-order_2019}. However, crystalline symmetries will play an important role in this work.
In addition, although a bulk-boundary correspondence can be established for fragility using twisted boundary conditions \cite{twisted_fragile}, this is difficult to implement in electronic systems so that a case-by-case analysis is still justified. }
In order to have a complete picture of fragile topological materials, a study
of fragile bands for all space groups in bulk materials within the framework
of TQC has been performed in
Ref. \cite{song_fragile_2020}, \changed{and high-throughput calculations have very recently classified two-dimensional materials attending to TQC and including fragile topology \cite{2DTQC}.}

\changed{Fragile topology has also 
been theoretically reported in photonic crystals and phonon spectra 
\cite{de_paz_engineering_2019,manes_fragile_2020,wei_fragile_2021,PhysRevB.110.045401}. One of its potential signatures}
is the presence of zero-dimensional corner states with corner charges, theoretically predicted in Refs. \cite{wieder_axion_2018,benalcazar_quantization_2019,ahn_failure_2019}. 
These corner charges have been theoretically reported in recent works studying
different mono- and few-layered materials with diverse rotational symmetries
\changed{\cite{schindler_fractional_2019-1,wang_higher-order_2019,kobayashi_fragile_2021,zeng_multiorbital_2021,kooi_bulk-corner_2021,PhysRevResearch.3.L032003,luo_fragile_2022,qian_c_2022,chen_fragile_2023,costa_connecting_2023,liu_magnetic_2023,nuñez_2024}}. 
Moreover, higher-order Fermi arcs have also been linked to fragile and higher-order topology
\cite{wieder_strong_2020}. 
From an experimental
point of view, fragile topology has been described in metamaterials
\changed{\cite{peri_experimental_2020,unal_PhysRevLett125,Jiang2021,Wu2024,Jiang2024} and quantum simulators \cite{Zhao2022}}, and Fermi arcs have been reported in some
transition-metal dichalcogenides 
\cite{rhodes_bulk_2017,di_sante_three-dimensional_2017,wu_three-dimensionality_2017}. 
Because corner charges are a shared feature between fragile and higher-order topology,
corner state measurements in real materials are often linked to the latter
\cite{schindler_higher-order_2018-1}. \changed{The potential applications of corner states as quantum dots have also been very recently highlighted, coining the concept of cornertronics \cite{Han2024}. }
  
It is therefore interesting to study the relation between higher-order and fragile
topology. In this direction, we construct a list of two-dimensional monolayers
described by space group (SG) $P\Bar{3}m1$ (164). 
\changed{Since transition-metal
dichalcogenides (TMDs) have received a great deal of attention as candidates for
layered materials with higher order 
\cite{wang_higher-order_2019,zeng_multiorbital_2021,qian_c_2022,costa_connecting_2023,liu_magnetic_2023,PhysRevMaterials.8.044203} and fragile topology \cite{wang_higher-order_2019,wieder_strong_2020},
we have focused on various TMDs, as well as on a number of non-TMDs with the
same space group.} We explore a total of 52 materials belonging to this SG, 
to find a set of experimentally suitable candidates.
We demonstrate that fragile topological bands are present in all 52 materials
in the region of energies near the Fermi level, and that these bands are \changed{related to}
the non-zero corner charges present in most of the materials and in $C_3$-symmetric
flakes.

This article is organized as follows: in Section II, the general properties of SG 164 and
the materials therein are discussed. In Section III, we compute a set of topological
invariants based on the symmetry eigenvalues of the system, particularized to the \changed{TMD} case
studies ZrSe$_2$, ZrS$_2$ and SnS$_2$. In Section IV, we theoretically predict the
corner charges linked to these materials and present the concept of partial corner
charge. Our results are consistent with first-principles corner states calculations
for triangular and hexagonal flakes in Section V.

\section{Computational methods and general background of space group 164 materials}

\begin{figure}[t!]
\includegraphics[width=\columnwidth,trim={10cm 5cm 22cm 6cm},clip]{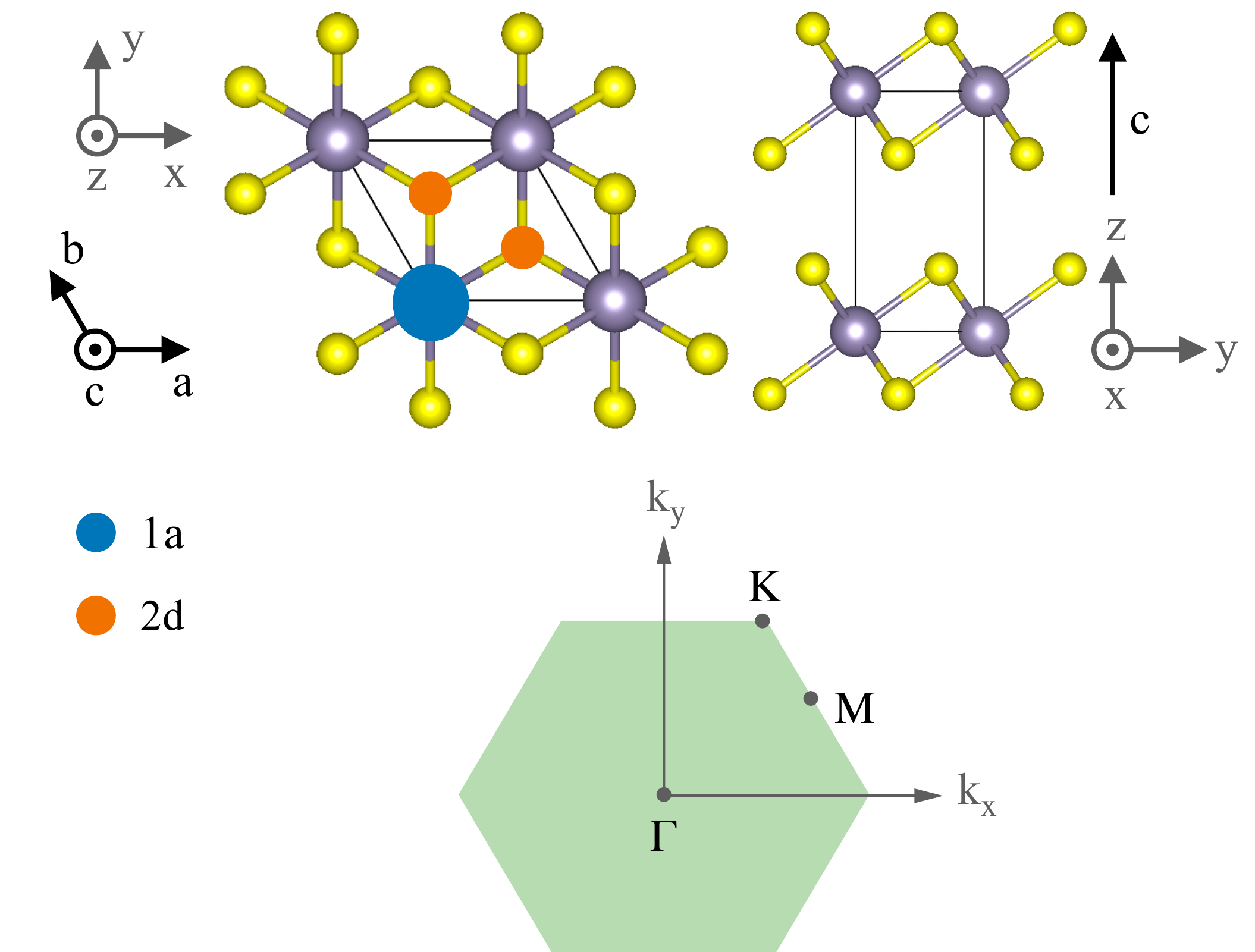}
\caption{Crystal structure of a unit cell belonging to SG 164: top view (upper left panel), and
side view (upper right panel) displaying two layers. The structure lattice vectors, as well as
the Cartesian axes, are indicated in black and gray, respectively. Blue and orange
highlighted atoms indicate the 1a and 2d WPs, respectively. Bottom panel: Brillouin zone of the
monolayer.}
\label{fig1}
\end{figure}

Electronic structure calculations for all materials were performed using the
Quantum ESPRESSO first-principles code
\cite{giannozzi_quantum_2009,giannozzi_advanced_2017} 
employing the generalized gradient approximation (GGA) and
Perdew–Burke-Ernzerhof (PBE) exchange-correlation functional. Spin-orbit coupling
was considered throughout the self-consistent calculations in all cases. A kinetic
energy wavefunction cutoff of 100 Ry and a $8\times 8\times 1$ Monkhorst-Pack reciprocal
space grid were used. All structures were relaxed until the forces were less than 0.001
eV/\AA. The initial crystal structures were obtained from the Computational 2D Materials
Database (C2DB) \cite{haastrup_computational_2018, gjerding_recent_2021} and checked
against the Topological Materials Database 
\cite{bradlyn_topological_2017-1,vergniory_complete_2019,vergniory_all_2022} for further
relaxation. 
Subsequently, a group theory analysis was performed using the IrRep code
\cite{iraola_irrep_2022}. For the flake calculations, the SIESTA code
\cite{soler_siesta_2002-2,garcia_siesta_2020} was used, including spin-orbit
coupling and considering the same GGA-PBE approximation.

SG 164 has inversion symmetry and includes rotational
symmetries $C_{3z}$ and $C_{2x}$, with a total of 12 symmetry operations. As shown
in Fig. \ref{fig1}, we focus on AB$_2$ compounds that display a layered
structure. The A atom, which is usually (but not necessarily) a transition metal, belongs
to the high-symmetry real space WP 1a, whereas the B atom is
described by the 2d WP. The spatial location of these positions within the unit cell
is highlighted in blue and orange, respectively, in Fig. \ref{fig1}. For the WPs,
as well as for the symmetry operations and the SG, we follow the
notation of the Bilbao Crystallographic Server
\cite{aroyo_bilbao_2006,aroyo_bilbao_2006-1,aroyo_crystallography_2011}.

We screen the C2DB by restricting the search to dynamically stable monolayers with SG 164 and AB$_2$ formula that possess a PBE band gap greater than 0.1 eV and a nonmagnetic ground state. Additionally, we also mention examples of semimetallic
systems that allow for a formal definition of a set of valence
bands separated from the upper conduction bands
by a non-absolute gap. A
complete list of materials identified in this work is presented in \changed{Table \ref{tab3}}. 

In the following sections we will focus on three illustrative materials of this collection that
have already been synthesized in monolayer form \cite{zhang_controlled_2015,manas-valero_raman_2016,ye_synthesis_2017}, namely, semiconductors ZrSe$_2$, ZrS$_2$, and SnS$_2$. The
semimetals ZrTe$_2$ and NiTe$_2$ are also briefly commented.
In addition to the aforementioned compounds, other materials encompassed in this work such as PtSe$_2$, SnI$_2$, 
\changed{PbI$_2$, PdS$_2$, PtS$_2$ and PtTe$_2$} are also experimentally available as monolayers \changed{\cite{wangMonolayerPtSe2New2015,sinhaAtomicStructureDefect2020,chen2DLayeredNoble2020,yuanDirectGrowthVan2021,liuEpitaxialGrowthMonolayer2021}. 
Note that the fragile topology database described in 
Ref. \cite{song_fragile_2020} may not apply here, since for monolayers the combination of EBRs is now restricted to the two-dimensional high-symmetry points of the Brillouin zone (BZ).}

\section{Band topology of $P\Bar{3}m1$ monolayers} 

\subsection{Topological quantum chemistry analysis}

\begin{figure*}[t!]
\includegraphics[width=0.9\textwidth,trim={0cm 6.5cm 0cm 0cm},clip]{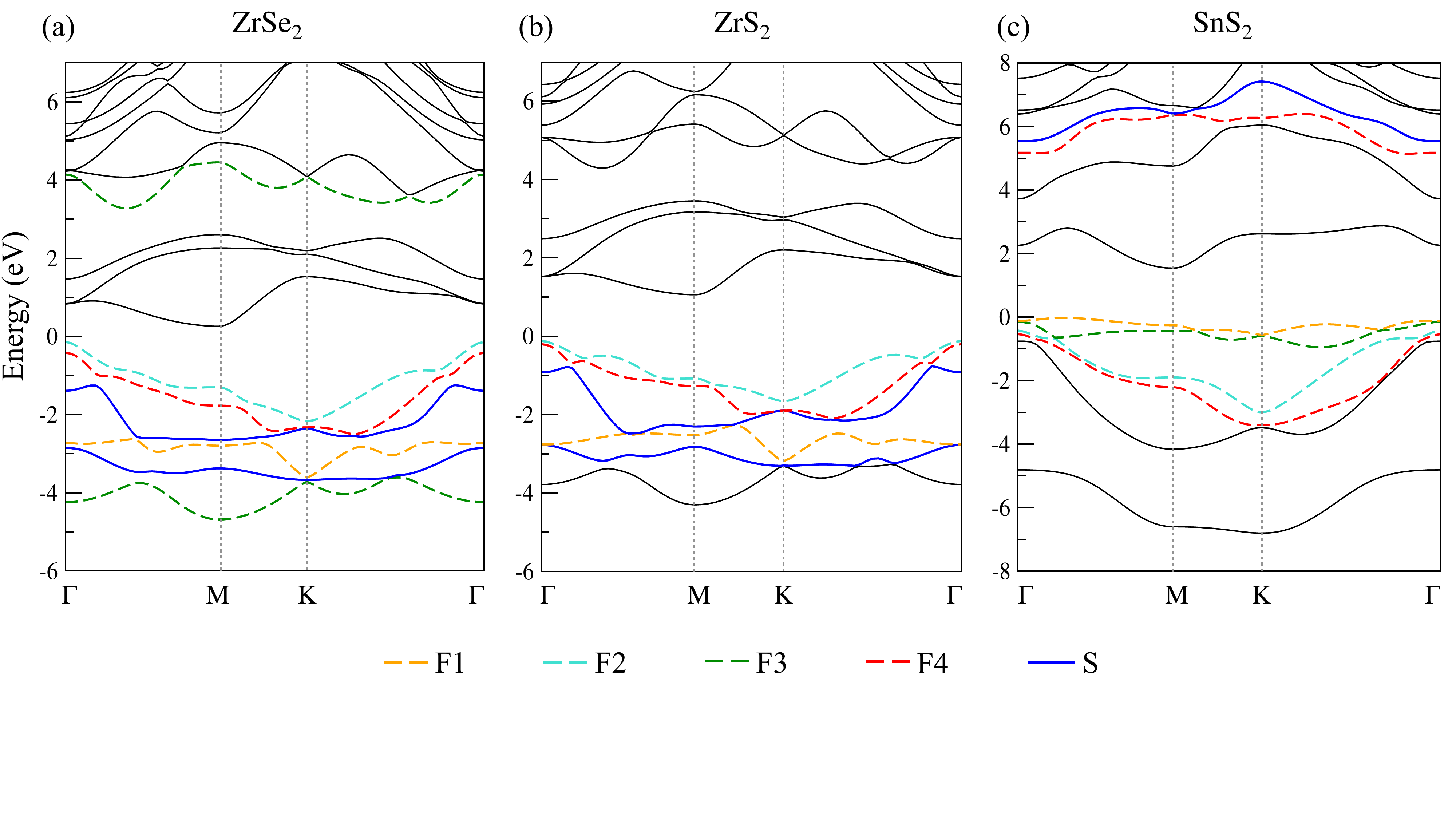}
\caption{Band structure of monolayer (a) ZrSe$_2$, (b) ZrS$_2$ and (c) SnS$_2$. Fragile bands are represented in colored dashed lines, and $\mathbb{Z}_2=1$ bands are highlighted in dark blue.}
\label{bands_tqc}
\end{figure*}

We perform a TQC analysis of all individual valence bands, as well as of the full valence
band manifold. 
\changed{The basic ingredient to start the topological description is to consider that time-reversal 
symmetry and spatial inversion are both preserved for all cases. This immediately implies
that all bands 
will be twofold degenerate along the entire
BZ. In addition, since spin-orbit is not negligible, double-space group irreducible
representations should be employed to characterize the bands at high symmetry points. 
In double SG 164, the highest degeneracy that a set of bands can have is two \cite{Elcoro2017}, so all doubly degenerate bands will always be isolated from each other. This allows for the topological classification not only of the valence band manifold as a whole, but also for the characterization of each twofold set of
bands. }
This basic set of bands is described by the labels of the irreps at high
symmetry points $\Gamma$, $M$ and $K$, which comprises what is known as a band
representation \cite{bradlyn_topological_2017-1}.

\begin{table}[h!]
    \centering
    \begin{tabular}{cccccc}
        & $\Gamma$  & $\mathrm{M}$ & $\mathrm{K}$ & $Q_{c,no\:inv}$ &  $Q_{c,inv}$\\ 
       \hline
       \rule{0pt}{3ex}EBR &$\Bar{\Gamma}_4\Bar{\Gamma}_5$ & $\Bar{\mathrm{M}}_3\Bar{\mathrm{M}}_4$ &$\Bar{\mathrm{K}}_4\Bar{\mathrm{K}}_5$ & 0& 0\\
       \color{orange}F1 &$\Bar{\Gamma}_4\Bar{\Gamma}_5$ & $\Bar{\mathrm{M}}_3\Bar{\mathrm{M}}_4$ &$\Bar{\mathrm{K}}_6$ &4/3 &2/3\\
       \color{blue}S &$\Bar{\Gamma}_4\Bar{\Gamma}_5$ & $\Bar{\mathrm{M}}_5\Bar{\mathrm{M}}_6$ & $\Bar{\mathrm{K}}_4\Bar{\mathrm{K}}_5$ &0 & 3/2\\
       \color{blue}S &$\Bar{\Gamma}_4\Bar{\Gamma}_5$ & $\Bar{\mathrm{M}}_5\Bar{\mathrm{M}}_6$ & $\Bar{\mathrm{K}}_6$ & 4/3 & 1/6\\
       \\
       \color{blue}S & $\Bar{\Gamma}_6\Bar{\Gamma}_7$ & $\Bar{\mathrm{M}}_3\Bar{\mathrm{M}}_4$ & $\Bar{\mathrm{K}}_4\Bar{\mathrm{K}}_5$ & 0 & 1/2\\
       \color{blue}S & $\Bar{\Gamma}_6\Bar{\Gamma}_7$ & $\Bar{\mathrm{M}}_3\Bar{\mathrm{M}}_4$ & $\Bar{\mathrm{K}}_6$ & 4/3 & 7/6 \\
       EBR & $\Bar{\Gamma}_6\Bar{\Gamma}_7$ & $\Bar{\mathrm{M}}_5\Bar{\mathrm{M}}_6$ & $\Bar{\mathrm{K}}_4\Bar{\mathrm{K}}_5$ &0 & 0\\
       \color{Cerulean}F2&$\Bar{\Gamma}_6\Bar{\Gamma}_7$ & $\Bar{\mathrm{M}}_5\Bar{\mathrm{M}}_6$ &$\Bar{\mathrm{K}}_6$ &4/3 & 2/3\\
       \\
       \color{DarkGreen}F3 &$\Bar{\Gamma}_8$ & $\Bar{\mathrm{M}}_3\Bar{\mathrm{M}}_4$ &$\Bar{\mathrm{K}}_4\Bar{\mathrm{K}}_5$ &2/3 & 4/3\\
       EBR &$\Bar{\Gamma}_8$ & $\Bar{\mathrm{M}}_3\Bar{\mathrm{M}}_4$ &$\Bar{\mathrm{K}}_6$ & 0 & 0\\
       \color{blue}S &$\Bar{\Gamma}_8$ & $\Bar{\mathrm{M}}_5\Bar{\mathrm{M}}_6$ & $\Bar{\mathrm{K}}_4\Bar{\mathrm{K}}_5$ & 2/3 & 5/6\\
       \color{blue}S &$\Bar{\Gamma}_8$ & $\Bar{\mathrm{M}}_5\Bar{\mathrm{M}}_6$ & $\Bar{\mathrm{K}}_6$ & 0 & 3/2\\
       \\
       \color{blue}S &$\Bar{\Gamma}_9$ & $\Bar{\mathrm{M}}_3\Bar{\mathrm{M}}_4$ & $\Bar{\mathrm{K}}_4\Bar{\mathrm{K}}_5$ &2/3 & 11/6 \\
       \color{blue}S & $\Bar{\Gamma}_9$ & $\Bar{\mathrm{M}}_3\Bar{\mathrm{M}}_4$ & $\Bar{\mathrm{K}}_6$ &0 & 1/2 \\
       \color{red}F4 &$\Bar{\Gamma}_9$ & $\Bar{\mathrm{M}}_5\Bar{\mathrm{M}}_6$ & $\Bar{\mathrm{K}}_4\Bar{\mathrm{K}}_5$ & 2/3 & 4/3 \\
       EBR &$\Bar{\Gamma}_9$ & $\Bar{\mathrm{M}}_5\Bar{\mathrm{M}}_6$ & $\Bar{\mathrm{K}}_6$ & 0 & 0 \\
    \end{tabular}
    \caption{Basic band representations in SG 164 monolayer, including trivial (labeled EBR), stable topological with $\mathbb{Z}_2=1$ (labeled S), and fragile (labeled F and numbered). Corner charges with and without inversion symmetry are gathered in the last two columns and described in Section IV. }
    \label{tab1}
\end{table}

\changed{Following the notation in the Bilbao Crystallographic Server, we list in Table \ref{tab1}
all possible sixteen band representations that are allowed by the compatibility
relations of the group.} These sixteen band
representations are classified as fragile, stable topological with
$\mathbb{Z}_2=1$, and trivial.  
If a set of bands corresponds to a combination of EBRs with integer coefficients and at
least one of them is negative, the set is fragile \cite{elcoro_application_2020}; if at
least one is fractional, the set has stable topology. Otherwise, if all coefficients
are positive integers, the set is generally said to be in an atomic limit and trivial
\cite{cano_band_2021}. To identify these phases, we follow the Smith decomposition
procedure \cite{elcoro_application_2020} as detailed in the Supporting Information (SI) \cite{SuppInfo}, and define a
strong topological invariant indicated by symmetry eigenvalues $\mathbb{Z}_2$ as
\begin{equation}
\mathbb{Z}_2=c_6\:\mathrm{mod\:2},
\end{equation}
where $c_6$ is the 6-th element of the $c$-vector, obtained from the multiplicities of the irreps of each band set and whose computation is detailed in the SI. 

\changed{We have checked that fragile and $\mathbb{Z}_2=1$ bands are present 
near the Fermi level  
in all reported materials. For instance, fragile band F4 can be decomposed into
EBRs as shown in Eq. \ref{eq_FB} below:
\begin{equation}
\begin{split}
\mathrm{F4}=&-{\Bar{E}_{1g}}@1a+{\Bar{E}_1}@2d.\\
\end{split}
\label{eq_FB}
\end{equation}
The negative coefficients confirm the fragile nature of the band, and are also present
in the rest of fragile bands in Table \ref{tab1}. 
In all cases, fragile bands are described in terms of EBRs coming from WPs 1a and 2d, which are the occupied sites in the unit cell. }
\changed{The EBR decomposition of all fragile bands F1 - F4, as well as of an exemplary strong topological band with $\mathbb{Z}_2=1$, are explicitly stated in the SI. Note that these representations are not unique as the general formulation for the symmetry data vector has several free parameters. These parameters are usually set to zero \cite{elcoro_application_2020}; however, we have checked that other parameter choices do not change the resulting topological nature of each symmetry data vector. }
A potential interplay between strong, 
$\mathbb{Z}_2=1$-indicated topological phases 
and fragile phases and its effect on 
corner states would be of great interest. 

\subsubsection*{Application to case studies}

In light of our preceding general results, we now classify the electronic bands in the representative case studies ZrSe$_2$, ZrS$_2$ and SnS$_2$. 
Fig. \ref{bands_tqc} (a) shows the band structure of monolayer ZrSe$_2$ near
the Fermi level, where a variety of topological bands, both stable and fragile, appear.
All types of identified fragile bands F1, F2, F3, and F4, shown in dashed lines and
different colors, are present in the valence set near the neutrality point. Stable
$\mathbb{Z}_2=1$ bands, highlighted in dark blue, also appear in the valence region and
have been \changed{broadly} labeled S for simplicity. If we extend the TQC analysis to the
entire valence band set, we find that ZrSe$_2$ is topologically trivial, \changed{as the total
set of irreps is expressed as a sum of different EBRs; the EBR decomposition of the full valence set is detailed in the SI}. 
This is consistent with
the intuitive notion of fragile bands: they are trivialized when combined with others.
However, the $\mathbb{Z}_2=1$ bands remain topological
upon addition; the only exception is the addition of an even number of
$\mathbb{Z}_2=1$ bands, since the $\mathbb{Z}_2$ index is defined modulo 2. We emphasize
that there are more topologically stable and fragile bands at lower energies, which
we do not show in the figure, but have been taken into consideration in order
to compute the topological properties of the valence band set. 

ZrSe$_2$ and ZrS$_2$ are semiconductors and 
share almost the same topological band
distribution for the uppermost valence bands, as seen in Fig. \ref{bands_tqc} (a)
and (b). In both cases, the first, second, and fourth valence bands are fragile, and the third and fifth valence bands are stable topological with
$\mathbb{Z}_2=1$. In fact, the band structure has an overall similar appearance
for both materials, as expected since they belong to a similar chemical
environment. However, in ZrS$_2$ there are no topological bands in the conduction
set near the Fermi level. As in ZrSe$_2$, all valence bands add up to a trivial
topological invariant. 

Figure \ref{bands_tqc} (c) displays an analogous band structure for the
semiconductor 
SnS$_2$. As in ZrSe$_2$, all fragile bands F1, F2, F3, and F4
found in Table \ref{tab1} are present in the figure. Interestingly, this Fig. does not include any $\mathbb{Z}_2=1$ bands; however, we have found such bands at lower energies. As in the previous cases, the whole
valence manifold is topologically trivial, \changed{as explicitly detailed in the SI}. 

For the rest of the monolayers studied, the results are similar. Additional figures highlighting fragile and stable bands for ZrTe$_2$ and NiTe$_2$ are found
in the SI. They both have a higher density of stable bands
but show fragile states near the Fermi level. Since ZrTe$_2$ and NiTe$_2$ are
semimetals, we have set the uppermost valence band to match previous works
\cite{bradlyn_topological_2017-1,vergniory_complete_2019,vergniory_all_2022}. ZrTe$_2$
is the only material studied with a topologically nontrivial valence band set, with $\mathbb{Z}_2=1$, \changed{in agreement with experimental reports of Dirac-like edge states \cite{tsipas_massless_2018}. 
In fact, }all the materials mentioned above (ZrSe$_2$, ZrS$_2$, SnS$_2$, ZrTe$_2$
and NiTe$_2$) are experimentally feasible in monolayer form.

\subsection{Symmetry-indicator invariants beyond topological quantum chemistry}

\changed{The former TQC analysis yields a broad range of materials with fragile and stable topological bands; however, we find that fragile bands become trivialized at the Fermi level. Moreover, even though a $\mathbb{Z}_2$ invariant is computed for stable topological bands, an invariant for fragile bands would also be useful to fully characterize the topological features of our system. }
A set of symmetry-indicator topological invariants, applicable to systems protected by \changed{crystalline (mainly rotational)} symmetries, has been introduced in Refs. \cite{schindler_fractional_2019-1,benalcazar_quantization_2019}. \changed{These invariants are computed in terms of certain symmetry eigenvalues of the system at selected high-symmetry points, and therefore characterize diverse topological phases, such as second-order HOTIs, fragile topology, or obstructed atomic limits \cite{schindler_fractional_2019-1}. }

\changed{In order to compute the corresponding symmetry-indicator invariants for our materials, we notice that all the studied monolayers belong to layer group $p\Bar{3}m1$. Following Ref. \cite{schindler_fractional_2019-1}, we focus on $C_3$ and inversion symmetries, present in layer group $p\Bar{3}m1$, in order to define a topological invariant for such a system in the presence of spin-orbit coupling:}
\begin{equation}
\nu_{inv}=(M^I_{-1},K^3_{-1}),
\end{equation}
where $M^I_{-1}$ is the difference between the number of states at $M$ and
$\Gamma$ points with inversion eigenvalue $-1$. Analogously, $K^3_{-1}$ is the
difference between the number of states at $K$ and $\Gamma$ points with $C_3$
eigenvalue $-1$. We will use $\nu_{inv}$ to characterize finite geometries that retain the monolayer point group ($\Bar{3}m$), and also explore the case of a
system in which inversion symmetry is absent, leaving $3m$ as the point group of
the structure.
For this configuration, $\nu_{no\:inv}$ is the appropriate invariant to describe the 
topological properties:
\begin{equation}
\nu_{no\:inv}=(K^3_{e^{i\pi /3}},K^3_{-1}),
\end{equation}
where $K^3_{e^{i\pi /3}}$ is the difference of number of states at $K$ and
$\Gamma$ with $C_3$ eigenvalue $e^{i\pi /3}$. 

\changed{The inversion-breaking scenario requires further elaboration because we are reducing 
the number of crystalline symmetries. }
To relate the topological classification
developed for SG 164 to this reduced case, we must study how the irreps of
this SG map to the SG without inversion, which is SG 156 ($P3m1$). 
As this is a subgroup of index 2 of SG 164, we can employ the subduction procedure, as 
explained in Ref. \cite{elcoro_application_2020}. The result of this map
can be extracted from the Bilbao Crystallographic Server, and for completeness, 
it
is also presented in the SI. 
\changed{Most importantly,} this map preserves the fragility of
the SG 164 bands. Moreover, bands with stable topology in SG 164 are mapped
to either EBRs (if the $K^3_{-1}$ invariant is zero) or fragile bands
(if the $K^3_{-1}$ invariant is nonzero) in SG 156. 
The prevalence of fragility in SG 156 is also confirmed by the Smith decomposition
procedure for this group, as it does not host stable $\mathbb{Z}_2$ phases that can
be diagnosed by symmetry indicators alone \changed{(there can, however, exist
other stable topological phases, as the system
belongs to the class AII of the Altland–Zirnbauer classification \cite{altland_nonstandard_1997})}.

\begin{table}
    \centering
    \begin{tabular}{cccccc}
    \toprule
    AB$_2$& $\#\mathrm{M}^I_{-1}$ & $\#\Gamma^I_{-1}$ &    $\#\mathrm{K}^3_{-1}$ & $\#\Gamma^3_{-1}$  & $\nu_{inv}$\\
    \midrule
    \midrule
    ZrSe$_2$ & 24 & 24 & 14 & 14 & $(0,0)$\\
    ZrS$_2$  & 14 & 14 & 8  & 6  & $(0,2)$\\
    SnS$_2$  & 8  & 8  & 10 & 8  & $(0,2)$\\
    \bottomrule\\[1pt]
    \toprule
    \rule{0pt}{3ex} AB$_2$ & $\#\mathrm{K}^3_{e^{i\pi /3}}$ & $\#\Gamma^3_{e^{i\pi /3}}$  &    $\#\mathrm{K}^3_{-1}$ & $\#\Gamma^3_{-1}$ & $\nu_{no\:inv}$
    \\[0.1cm]
    \midrule
    \midrule
    ZrSe$_2$ & 15 & 15 & 14 & 14 & $(0,0)$\\
    ZrS$_2$  & 8 & 9 & 8  & 6  & $(-1,2)$\\
    SnS$_2$  & 8 & 9 & 10 & 8 & $(-1,2)$\\
    \bottomrule
    \end{tabular}
    \caption{Symmetry indicator invariants related to fractional corner charges in SG 164 monolayer unit cell with (top part) and without (bottom part) inversion symmetry.}
    \label{tab3}
\end{table}
\begin{figure*}[ht!]
\includegraphics[width=0.9\textwidth,trim={0cm 5cm 0cm 0cm},clip]{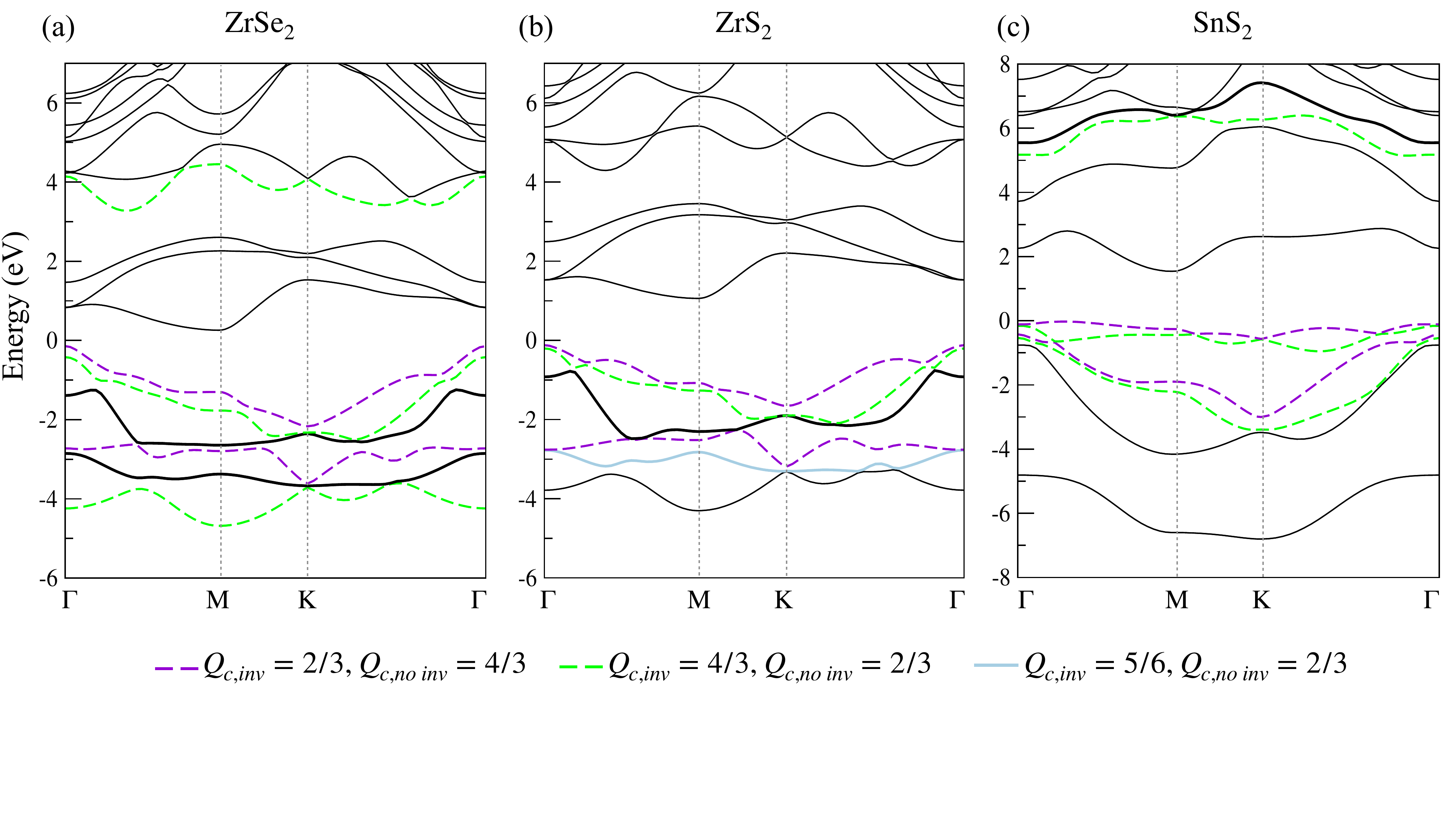}
\caption{Band structure of monolayer (a) ZrSe$_2$, (b) ZrS$_2$ and (c) SnS$_2$ with bands with nonzero partial corner charges colored in green and purple, respectively. Bands with ill-defined corner charges are colored gray. Fragile bands are represented with dashed lines, and stable topological bands are highlighted with thick lines.}
\label{bands_qc}
\end{figure*}

\subsubsection*{Application to case studies}

Table II collects \changed{both symmetry indicators $\nu_{inv}$ and $\nu_{no\:inv}$} for the valence band set
of the selected compounds ZrSe$_2$, ZrS$_2$, and SnS$_2$. \changed{Although the TQC analysis presented in the previous section yielded a trivial valence band set for all three compounds, nontrivial symmetry-indicator invariants are obtained for all cases except ZrSe$_2$. This is expected, since symmetry-indicator invariants successfully characterize topological phases whose bands at the Fermi level are an apparently trivial sum of EBRs, such as fragile topology and obstructed atomic limits. If follows that symmetry indicators are especially useful in our systems. Crucially, a nonzero $\nu_{inv}$ anticipates the presence of fractional corner charges in $C_3+I$-symmetric materials under open boundary conditions, whereas a nonzero $\nu_{no,\:inv}$ indicates a similar behavior for $C_3$-symmetric systems \cite{benalcazar_quantization_2019,schindler_fractional_2019-1}}. 

On a different note, symmetry-indicator invariants are also related to the Wilson loop spectra of the bands
\cite{schindler_fractional_2019-1}, which, due to the presence of crystalline
symmetries, can be computed using the corresponding eigenvalues. \changed{As a general result for all the structures studied, we find that all fragile bands present a zero $M^{I}_{-1}$ invariant, which indicates a nonwinding Wilson loop along the
$\Gamma-M$ direction. This is the usual path for the characterization of the $\mathbb{Z}_2$ topology and indicates that fragile bands will show a trivial character along that direction even when their WL spectra are calculated in isolation from other bands. 
Nevertheless, as has been made patent in previous works \cite{bradlyn_disconnected_2019,bouhon_PhysRevB.100.195135,kobayashi_fragile_2021,luo_fragile_2022,henke_PhysRevB104}, fragility 
can be captured by other different paths, a fact that the nonzero values of
$ K^3_{e^{i\pi /3}}$ and $K^3_{-1}$ corroborate. These invariants are related
to the $\Gamma-K$ path, which implies that fragile bands have nontrivial features that will show in responses related to these indicators. }

\section{Corner charge analysis}

In an open system,
fractional corner charges are a signature of topology stemming from \changed{a
filling anomaly, which arises at charge neutrality when enforcing certain crystalline symmetries in topological systems \cite{schindler_fractional_2019-1,benalcazar_quantization_2019,lee2020,khalaf_boundary-obstructed_2021,hitomi2021}.} These systems
can be trivial under periodic boundary conditions, but not under open conditions, \changed{where robust, fractionally-filled corner states appear \cite{schindler_fractional_2019-1,wieder_strong_2020}}. 
In particular, in the presence of $C_3$ and
inversion symmetry, the corner charge $Q_{c,inv}$ hosted at each $C_3$-symmetric sector of a flake is defined in terms of the symmetry-indicator invariants defined in Section III as \cite{schindler_fractional_2019-1}
\begin{equation}
Q_{c,inv}=-\frac{1}{4}M_{-1}^I-\frac{1}{3}K_{-1}^{3} \:\mathrm{mod\:2}.
\label{Q_c_inv}
\end{equation}
The allowed charge values for the three equal threefold symmetric corners, besides
zero, are \changed{$1/3$, $2/3$, $1$, $4/3$, or $5/3$}. In the absence of
inversion, the corner charge $Q_{c,no\:inv}$ reads
\begin{equation}
Q_{c,no\:inv}=\frac{2}{3}K_{e^{i\pi /3}}^{3}+\frac{2}{3}K_{-1}^{3} \:\mathrm{mod\:2}.
\label{Q_c_noinv}
\end{equation}
This implies three equal corner charges with values \changed{$2/3$ or $4/3$}. These formulas hold in the presence of spin-orbit coupling. \changed{For both the inversion-preserving and the inversion-breaking cases, the presence of spin-orbit coupling and time-reversal symmetry ensures that all states are at least twofold degenerate due to the Kramers' theorem, which leads to the aforementioned allowed values for $Q_{c,inv}$ and $Q_{c,no\:inv}$ \cite{benalcazar_quantization_2019,schindler_fractional_2019-1}}.

\changed{It is worthwhile to emphasize the non-uniqueness of the corner charge value \cite{benalcazar_quantization_2019}, since the filling anomaly at the neutrality point can be overcome by either filling or depleting the Fermi level. In the following, we will work in units of the fundamental charge of the electron, and therefore assume that a corner charge of $n/N$ indicates that each $N$-symmetric sector of the system would need   $n$ additional electrons for the states at the neutrality point to be completely filled. This convention has also been explicitly followed in previous works \cite{Sdequist2022}.}

In this way, it is straightforward to predict the presence of corner charges $Q_{c,inv}$ and $Q_{c,no\:inv}$ in candidate materials from standard first-principles calculations. \changed{The existence of corner charges, that could be observed via STM measurements, is key for the experimental identification of novel types of topology such as HOTIs and fragile systems. The relevance of corner charges is further highlighted since, as discussed in the Introduction, realizations of fragile topology are elusive, but nonzero $Q_{c,inv}$ or $Q_{c,no\:inv}$ would confirm that the fragility of these systems could be overcome in an open boundary setting}. 

\changed{However, as already commented in Refs. \cite{benalcazar_quantization_2019,schindler_fractional_2019-1}, symmetry-indicator invariants and corner charges are a shared property between second-order HOTIs, fragile systems, and obstructed atomic limits. In fact, besides stable and fragile bands, some SG 164 monolayers also show obstructed atomic insulator  behavior \cite{Sheng2024,OOAI}. Symmetry-indicator invariants $\nu_{c,inv}$ and $\nu_{c,no\:inv}$ do not distinguish between the aforementioned topological phases, since they are the same for all of them. Nevertheless, the presence of corner charges is mostly linked to HOTIs in the literature. Therefore,} we aim to know whether the potential (well-defined) corner states are exclusively related to fragile topology \changed{in SG 164 monolayers} or not. To this end, we follow two equivalent approaches. First, \changed{we compute a partial corner charge, which is the application of the above equations for $Q_{c,inv}$ and $Q_{c,no\:inv}$ to each set of doubly degenerate bands. Thus, we can identify which bands contribute to the total
corner charge, and from the TQC analysis of Section III A, the type of topology they host. Second, we obtain the total value of the corner charges for the complete valence band set, which results from the addition (modulo 2) of the partial corner charges}.

\subsubsection*{Application to case studies}

We first present the result of the partial corner charge approach in panels (a), (b), and (c) of Figure
3 for ZrSe$_2$, ZrS$_2$ and SnS$_2$, respectively. All three materials present fragile bands with well-defined nonzero corner charges
$Q_{c,inv}$ and $Q_{c,no\:inv}$ near the Fermi level, which could be accessed
individually in principle by doping processes. Analogous Figs. for semimetals ZrTe$_2$ and NiTe$_2$ can be found in the SI. 

In Table \ref{tab1} we present a summary of the computed partial corner charges
for the sixteen types of basic band representations. It is readily observed
that EBRs do not carry any partial corner charges, and that 
all the nonzero values stem from fragile and topologically stable bands. In
particular, in the case with no inversion, as \changed{from Section III B} only fragile bands are 
present in SG 156, a nonzero total corner charge is 
exclusively associated with the presence of uncompensated (fragile) partial corner
charges. 

On the other hand,
when inversion is present, 
\changed{fragile bands individually have a 
well-defined fractional corner charge according to the definition
\cite{benalcazar_quantization_2019,schindler_fractional_2019-1}, whereas stable bands
present corner charges that are not among the allowed values of Eq. 5 and thus could not give rise to the corresponding filling anomaly.} This is
the case, for example, of the fifth valence band of ZrS$_2$ in Figure
3 (b). \changed{However,} 
the sum of an even number of stable topological bands (which will trivialize the $\mathbb{Z}_2$ 
invariant) results in a 
direct sum of EBRs and fragile bands, with a well-defined joint partial corner charge. 
For instance, the sum of the fifth and eighth stable bands in Table \ref{tab1}, both present
in ZrS$_2$, is
\begin{equation}
\begin{split}
\mathrm{S5}+\mathrm{S8}&=(\Bar{\Gamma}_9\oplus \Bar{\mathrm{M}}_3\Bar{\mathrm{M}}_4\oplus \Bar{\mathrm{K}}_6) \oplus (\Bar{\Gamma}_8\oplus \Bar{\mathrm{M}}_5\Bar{\mathrm{M}}_6\oplus \Bar{\mathrm{K}}_4\Bar{\mathrm{K}}_5)\\
&= (\Bar{\Gamma}_9\oplus\Bar{\mathrm{M}}_5\Bar{\mathrm{M}}_6\oplus\Bar{\mathrm{K}}_6) \oplus 
(\Bar{\Gamma}_8\oplus\Bar{\mathrm{M}}_3\Bar{\mathrm{M}}_4\oplus\Bar{\mathrm{K}}_4\Bar{\mathrm{K}}_5)\\
&=\mathrm{EBR}+\mathrm{F3}.
\end{split}
\end{equation} 

\begin{table}[h!]
\centering
\begin{tabular}{c|ccccc}
\toprule
A & \multicolumn{4}{c}{B = Chalcogen} \\
\midrule
\midrule
      Zr &  \textcolor{Green}{ZrO$_2$}& \color{DarkGreen}{\textbf{ZrS$_2$}}&  \textbf{ZrSe$_2$}& \textcolor{Gray}{\textbf{ZrTe$_2$}}\\
      Sn &  \textcolor{Green}{SnO$_2$}& \textcolor{Green}{\textbf{SnS$_2$}} & SnSe$_2$\\
      Ni &  \textcolor{Green}{NiO$_2$}& \textcolor{Green}{NiS$_2$}& & \textbf{NiTe$_2$}\\
      Pt &  \textcolor{Green}{PtO$_2$}& \textcolor{Green}{PtS$_2$}&PtSe$_2$&PtTe$_2$\\      
      Hf &  \textcolor{Green}{HfO$_2$}& \textcolor{Green}{HfS$_2$}&HfSe$_2$\\
      Pd &  \textcolor{Green}{PdO$_2$}& \textcolor{Green}{PdS$_2$}&PdSe$_2$\\
      Si & & & SiSe$_2$ \\
      Pb &  \textcolor{Green}{PbO$_2$}& \textcolor{Green}{PbS$_2$}\\
      Ge &  \color{DarkGreen}{GeO$_2$}& \color{DarkGreen}{GeS$_2$}\\ [2pt]
      \toprule
      \rule{0pt}{3ex}A & \multicolumn{4}{c}{B = Halogen}
      \\[0.1cm]
      \midrule
      \midrule
      Cd &  \textcolor{Green}{CdI$_2$} & \textcolor{Green}{CdBr$_2$} & \textcolor{Green}{CdCl$_2$}& \textcolor{Green}{CdF$_2$} \\
      Zn &  \textcolor{Green}{ZnI$_2$} & \textcolor{Green}{ZnBr$_2$}& \textcolor{Green}{ZnCl$_2$}& \textcolor{Green}{ZnF$_2$}&   \\
      Pb &  \textcolor{Green}{PbI$_2$} & \textcolor{Green}{PbBr$_2$} & \textcolor{Green}{PbCl$_2$} \\    
      Ru &  \textcolor{Green}{RuI$_2$} &  \textcolor{Green}{RuBr$_2$}& \textcolor{Green}{RuCl$_2$} \\
      Hg &  & & & \textcolor{Green}{HgF$_2$}\\
      Mg &  \textcolor{Green}{MgI$_2$}&  \textcolor{Green}{MgBr$_2$}\\
      Ca &  \textcolor{Green}{CaI$_2$} & \textcolor{Green}{CaBr$_2$} \\
      Sr &  \textcolor{Green}{SrI$_2$}& \textcolor{Green}{SrBr$_2$}\\
      Ba &  \textcolor{Green}{BaI$_2$}& \textcolor{Green}{BaBr$_2$}\\
      Os &  \textcolor{Green}{OsI$_2$}& \textcolor{Green}{OsBr$_2$}\\  
      Sn &  \color{DarkGreen}{SnI$_2$}\\
      Ge &  \color{DarkGreen}{GeI$_2$}\\[2pt]
      \bottomrule
    \end{tabular}
    \caption{List of all the materials studied in the database, \changed{sorted by the elements A and B in their formula AB$_2$. The exemplary materials discussed in the main text and the SI are highlighted in bold text.} Compounds highlighted in green have $Q_{c,inv}=4/3$ and $Q_{c,no\:inv}=2/3$, in gray $Q_{c,inv}$ is not well defined, and the rest have no corner charges. }
    \label{tab3}
\end{table}

This set 
has a corner charge $Q_{c,inv}=4/3$, which is a well-defined value and the same as that of fragile band F3. We have checked that this holds for all pairwise additions of strong bands in Table \ref{tab1}. Moreover, even when partial corner charges stemming from fragile bands are trivialized mod $2$ and the only contribution to $Q_{c,inv}$ is that of stable bands, these stable bands always come in pairs so that their joint contribution equals a sum of EBRs and fragile bands. Detailed calculations that support this statement can be found in the SI for several materials.

\begin{figure*}[t!]
\includegraphics[width=0.9\textwidth,trim={0cm 3.5cm 0 0.5cm},clip]{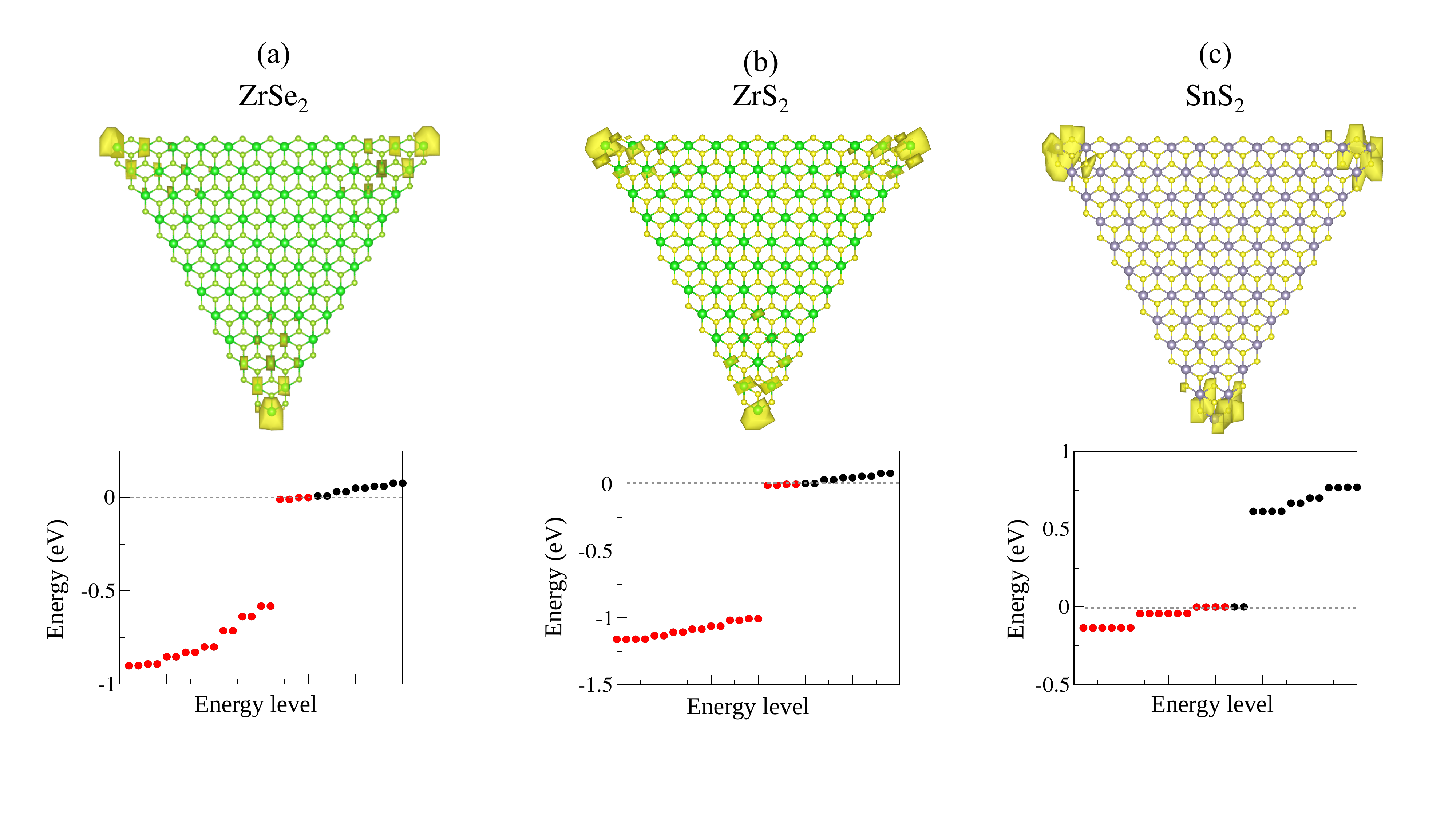}
\caption{LDOS (top panel) and eigenvalue spectra (bottom panel) for the (a) ZrSe$_2$,
(b) ZrS$_2$ and (c) SnS$_2$ triangular flakes.}
\label{flakestri}
\end{figure*}

It follows that the origin of the corner charges in this SG
is \changed{related} to fragile topology, even when $\mathbb{Z}_2=1$ bands
are considered. Even when the whole valence band manifold is trivialized \changed{from a TQC perspective, corner states can appear}. 

We have verified that the sum of partial charges equals the corner charge obtained
by applying Eqs. \ref{Q_c_inv} and \ref{Q_c_noinv} for the whole set of
valence bands for the five materials highlighted in this work. The results for all the valence bands for each material are gathered
in \changed{Table \ref{tab3}}
. For all materials in our database, well-defined corner charges
$Q_{c,inv}=4/3$ or $Q_{c,inv}=0$, and $Q_{c,no\:inv}=2/3$ or $Q_{c,no\:inv}=0$ are
found, except for ZrTe$_2$, that has an ill-defined $Q_{c,inv}=5/6$. This is
consistent with the fact that ZrTe$_2$ is stable topological with 
$\mathbb{Z}_2=1$ \changed{and hosts Dirac cones, which hamper the presence of a filling anomaly at the neutrality point}. Moreover, over
80\% of the samples have non-zero corner charges.  

\changed{However, we have found that all halogen-based materials show a mismatch between the computed $Q_{c,inv}=4/3$, $Q_{c,no\:inv}=2/3$ and the actual filling anomaly inferred from their flake eigenvalue spectra, as demonstrated in the SI for MgBr$_2$. Namely, by counting the number of filled states at the neutrality point, we find that flakes involving halogens should have $Q_{c,inv}=5/3$ and $Q_{c,no\:inv}=4/3$.} 

\changed{Even though chalcogen- and halogen-based SG 164 monolayers have identical crystalline symmetries, they do not host the same number of filled states at the neutrality point. This difference in the filling anomaly value stems from the fact that the 2d WP in chalcogen-based materials has no atomic support in real space, whereas for halogen-based systems it has partial support \cite{OOAI}. In fact, the fractional corner charge of a system can be foreseen from symmetry arguments based on the real-space distribution of Wannier functions \cite{benalcazar_electric,benalcazar_quantization_2019,schindler_fractional_2019-1}, and SG 164 monolayers also host a topological orbital-obstructed atomic insulator phase \cite{Sheng2024,OOAI}. 
It follows that, unlike stable topological bands (whose effect on corner charges could be traced back to fragile bands), orbital-obstructed topology plays a major role along with fragile topology in the emergence of corner states in SG 164 monolayers. The trivialization of the fragile bands at the Fermi level, where obstructed atomic insulators are diagnosed, impedes to study both topological phases separately. Since for chalcogen-based materials fractional corner charges can be determined solely from crystalline symmetry arguments and a more detailed study of the number and disposition of the valence electrons is needed for halogen systems, in the following we will focus on the former and leave the latter for a different study \cite{OOAI}. Overall, for materials with no atomic support at the 2d WP, the arguments presented in this work are valid.}

Besides the three highlighted systems, other materials with nonzero
fractional charges are \changed{PdS$_2$, SnS$_2$ or PtS$_2$}, which are synthesized in monolayer form and constitute interesting
experimental candidates for STM measurements of corner charges.

\section{Corner states in flakes}

\begin{figure*}[t!]
\includegraphics[width=0.9\textwidth,trim={0cm 3.5cm 0 0.5cm},clip]{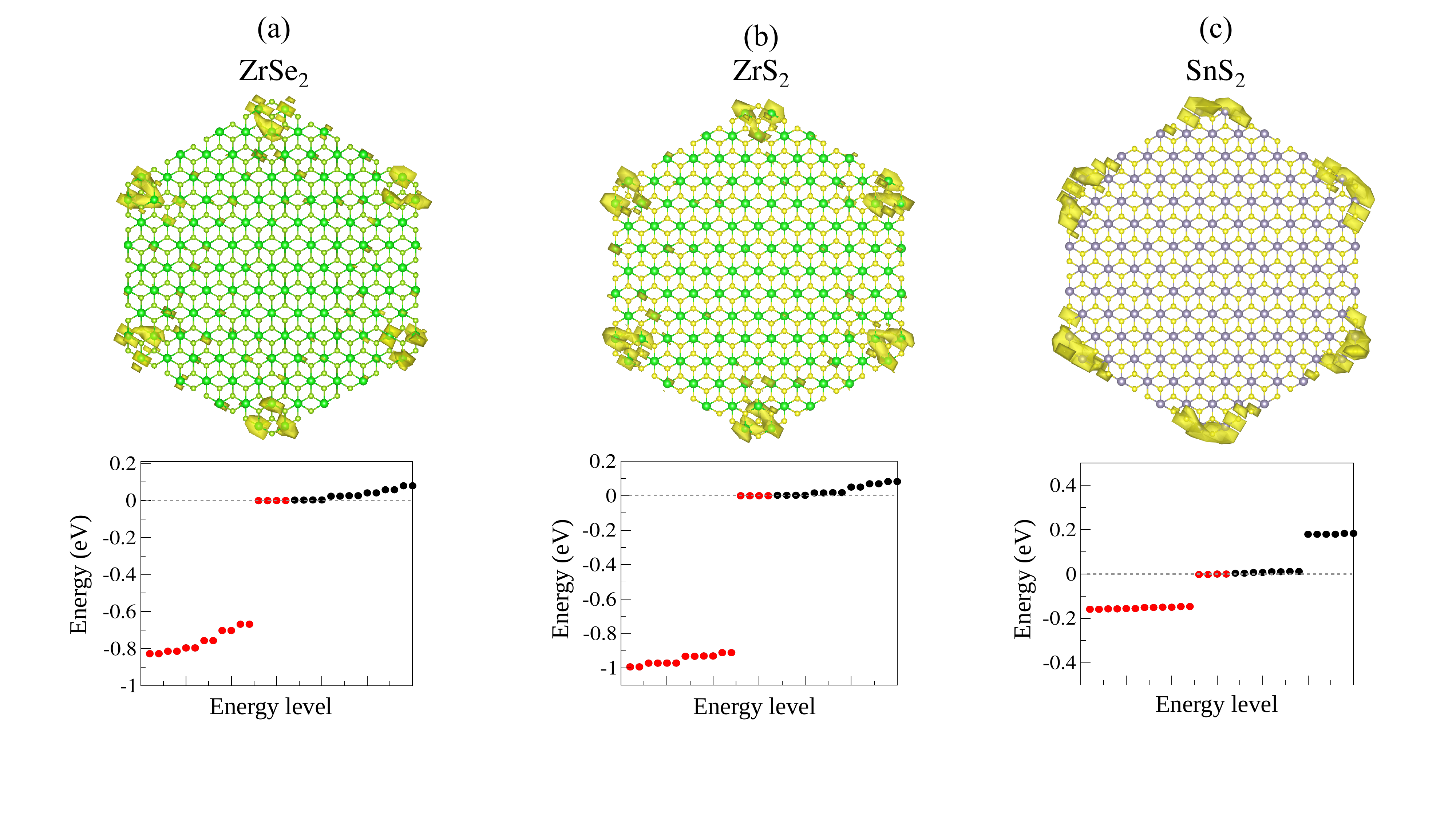}
\caption{LDOS (top panel) and eigenvalue spectra (bottom panel) for the (a) ZrSe$_2$, (b) ZrS$_2$ and (c) SnS$_2$ hexagonal flakes. }
\label{flakeshex}
\end{figure*}

In contrast to most materials hosting fragile topology, SG 164 does not have
$C_{2z}$ symmetry, so our fragile bands must be protected by other symmetry operations.
Following our previous reasoning, 
we propose two flake geometries, a
hexagonal one with $C_3$ symmetry and inversion, and a triangular one with $C_3$ but
without inversion. The latter is the most simple approach since the 
triangular flake breaks not
only the inversion but also the $C_2$ symmetries, having point group symmetry $3m$.
For its part, the hexagonal flake preserves all monolayer symmetries, with point
group symmetry $\Bar{3}m$.

For the triangular flake, a high electronic localization is observed via first-principles local
density of states (LDOS) calculations at the three corners of Fig. \ref{flakestri} for the
selected materials ZrSe$_2$, ZrS$_2$, and SnS$_2$. \changed{Since a band gap is necessary for well-defined corner charges, semimetals ZrTe$_2$ and NiTe$_2$ are not considered here}. 

In the bottom panel \changed{of Fig. \ref{flakestri}, the flake eigenvalues depict a sixfold}
degeneracy at the Fermi level. Note that the states are not exactly localized at the
Fermi level due to the absence of chiral symmetry \cite{luo_fragile_2022}. DFT simulations also take
into account potential interactions between corner states \cite{liu_two-dimensional_2019}, which contribute to the elimination of degeneracy. \changed{This size effect is reduced by
increasing the flake size; in fact, the number of degenerate states at the neutrality point can change depending on the system size, although the value of filled states is left unchanged (see SI). Such degeneracy is best observed the larger the band gap is, as verified by considering that SnS$_2$ has the largest gap of all three materials, and also comparing with Fig. 3 in the SI for MgBr$_2$ (with a gap over three times larger than SnS$_2$).}

To fill this \changed{sixfold} degeneracy, two additional
electrons are needed for each material; therefore, each corner of the triangle should
contribute $2/3$ \changed{(in units of the electron charge). From Table \ref{tab3}, both ZrS$_2$ and SnS$_2$ have a corner
charge without inversion $Q_{c,no\:inv}=2/3$, in agreement with the value obtained by inspection of the eigenvalue spectra. The
only exception is ZrSe$_2$, whose eigenvalue spectra yields $Q_{c,no\:inv}=2/3$ but has a computed $Q_{c,no\:inv}=0$ from Table \ref{tab3}. This is due to the trivialization of partial corner charges by the extra low-lying valence bands that are considered within our DFT framework for Se-based alloys, and that are absent in other different DFT approaches \cite{2DTQC}.}

Fig. \ref{flakeshex} shows the LDOS and energy spectra of hexagonal flakes \changed{with $C_3+I$ symmetry. As in Fig. \ref{flakestri}, there are four filled states at the neutrality point; however, now degeneracies are increased due to the presence of inversion symmetry. Namely, the spectrum is 12-fold degenerate at the Fermi level; this is best seen for flakes with larger band gaps whose edge states sit higher in energy, such as SnS$_2$ in panel (c). In fact, panels (a), (b) and (c) are ordered by increasing band gap.
In order to fill the Fermi level, 8 extra electrons would be needed, yielding $Q_{c,inv}=4/3$ per corner of the flake. This value is in agreement with our previous calculations of the corner charge in Table \ref{tab3}. Additional hexagonal flake figures are shown in the SI for smaller systems, emphasizing the need for sizable flakes in order to obtain robust degeneracies. }

\section{Conclusions}

SG 164 monolayer systems span a great variety of structures, which motivates a
thorough search for nontrivial topological properties in this family. 
A comprehensive selection of compounds with AB$_2$ formula was
surveyed, and the case studies ZrSe$_2$, ZrS$_2$ and SnS$_2$ were selected for a
detailed description.
Our general results indicate the presence of stable ($\mathbb{Z}_2=1$) and
fragile topological bands. \changed{Thus, SG 164 is an interesting platform to study
the interplay between fragile and stable topology and its experimental implications, and where other topological phases such as orbital-obstructed atomic insulator behavior have also been reported.} 

Using first-principles calculations, we have analyzed the symmetry eigenvalues of
the topologically nontrivial bands in these materials and computed their corner
charges and related invariants. We have found nonzero topological indices for more
than 80\% of the screened materials. \changed{A detailed analysis of partial corner
charges, computed for each band, has proven fundamental in clarifying the role
of stable topological and fragile bands. }
Whereas these 
fractional charges are
usually related to higher-order topology, we show that, in \changed{many} SG 164 monolayers, corner
charges \changed{are closely related} to fragile bands. Moreover, via irrep subduction to the SG 156,
we demonstrate that the nontrivial character of the fragility-induced 
topological invariants must be preserved in configurations where inversion is
explicitly broken. \changed{Halogen-based monolayers are an exception, since the predicted corner charges are not in agreement with subsequent flake calculations. For these materials, a real-space analysis following an orbital-obstructed study can correctly foresee corner charges, and fragile topology plays a secondary role.}

These findings were therefore tested \changed{mainly on non-halogen compounds, and} using triangular and hexagonal geometries. For
triangular flakes, a filling anomaly was identified for the three exemplary
materials, resulting in a corner charge valued $2/3$. Regarding
hexagonal structures, \changed{we find a corner charge of $4/3$. Both values agree with our previous analytical calculations.} 

It can also be noted that as the nontrivial results depend on the $3m1$
symmetry, materials with a bulk structure described by SG 156 are likely to
present the same effects. Examples of this are Janus TMDs, which are seen as
an attractive platform for extending and applying the results achieved in this
work. In summary, a direct link was established between $C_3$ symmetry-protected corner
charges and fragile topology in a wide set of two-dimensional materials. The
universality of our findings among SG 164, and possibly other SGs sharing the
relevant threefold symmetry, encourages the experimental realization of
responses bound to fragile topology in real materials.

\section*{Acknowledgements}

The authors thank J. D. Correa for helpful discussions. We acknowledge the
financial support of the Agencia Estatal de Investigación of Spain under grant
PID2022-136285NB-C31 and from the Comunidad de Madrid through the (MAD2D-CM)-UCM5
project, funded by the Recovery, Transformation and Resilience Plan, and
by NextGenerationEU from the European Union. O.A.G acknowledges the support of grant PRE2019-088874 funded by
MCIN/AEI/10.13039/501100011033 and by “ESF
Investing in your future”\changed{, and of European Union NextGenerationEU/PRTR project Consolidación Investigadora CNS2022-136025, as well as Instituto de Ciencia de Materiales de Madrid (ICMM-CSIC) and Universidad Complutense de Madrid, where this project was initiated}. S.B. acknowledges the support of the Postdoctoral
Grant from the Universidad Técnica Federico Santa María. M. P. 
acknowledges the financial support of Chilean FONDECYT by grant 1211913. \changed{Finally, we thank the Centro de Supercomputación de Galicia, CESGA, (www.cesga.es, Santiago de Compostela, Spain) and Supercomputación Castilla y León (SCAYLE) for providing access to their supercomputing facilities. }

\section*{Supporting Information}
The Supporting Information is available at \url{http://link.aps.org/supplemental/10.1103/qcr8-jxvn}. 

It includes topological quantum chemistry analysis of SG 164 monolayers, as well as additional figures for topological semimetals ZrTe$_2$ and NiTe$_2$ and other flake figures. The irreducible representations and symmetry operations at high symmetry points of all the valence bands of the 52 materials studied using Quantum ESPRESSO and the IrRep code are available and can be downloaded as a zip file.

\bibliography{ref}
\end{document}